\documentclass[final]{svjour2}
\usepackage{graphicx}
\usepackage{wrapfig}
\usepackage{rotating}
\usepackage{amssymb}
\usepackage{bm}
\usepackage{mathptmx}
\usepackage[numbers]{natbib}
\makeatletter
\journalname{Journal of Low Temperature Physics}

\bibpunct{}{}{,}{s}{}{,}

\begin{document}

\newcommand{\hdblarrow}{H\makebox[0.9ex][l]{$\downdownarrows$}-}
\title{Counterflow Quantum Turbulence and the Instability in Two-component Bose-Einstein Condensates}

\author{S. Ishino \and H. Takeuchi \and M. Tsubota }

\institute{Department of Physics, Osaka City University,Sumiyoshi-Ku, Osaka 558-8585, Japan\\
Tel.:+81-6-6605-2501\\ Fax:+81-6-6605-2522\\
\email{ishino@sci.osaka-cu.ac.jp}
}

\date{XX.06.2010}

\maketitle


\begin{abstract}

 We theoretically study the nonlinear dynamics of the instability of counter-superflow in two miscible Bose-Einstein condensates.
 The condensates become unstable when the relative velocity exceeds a critical value, which is called counter-superflow instability.
 We reveal that the counter-superflow instability can lead to quantum turbulence by numerically solving the coupled Gross-Pitaevskii equations.
 The modes amplified by the instability grow into solitons and decay into quantized vortices.
 Eventually, the vortices become tangled and quantum turbulence of two superfluids.
 We show that this process may occur in experiments by investigating the dynamics in a 2D trapped system.

\end{abstract}

\section{Introduction}

Quantum turbulence (QT) is currently one of the most important topics in low temperature physics \cite{PLTP}.
QT has been historically studied in superfluid $^4$He and $^3$He, while QT is now numerically studied in atomic Bose-Einstein condensates (BECs) \cite{Kobayashi07} and experimentally realized recently \cite{Henn09}.
Atomic BECs have several advantages over superfluid helium.
The most important point is that modern optical techniques enable us to easily control the condensate and directly visualize quantized vortices.

Old studies on QT in superfluid $^4$He focused primarily on thermal counterflow, in which the normal fluid and superfluid flow in opposite directions driven by an injected heat current. 
In this work we consider a similar case in two-component BECs.
Two-component BECs are known to create various exotic structure of quantized vortices\cite{Kasamatsu05} and cause some characteristic hydrodynamic instability such as Kelvin-Helmholtz instability \cite{Takeuchi10} and Rayleigh-Taylor instability \cite{Sasaki09}.  
Hence QT and the instability towards QT in two-component BECs should propose novel dynamical phenomena and open the new research field in cold atoms.
For example, the energy spectra of QT in a single-component BEC is known to obey the Kolmogorov law that is the most important statistical law in turbulence \cite{Kobayashi05}, but it is not so trivial what happens to QT of two-component BECs. 

In this work we study numerically QT in counterflow of two-component miscible BECs for the first time (Fig. 1). 
Law {\it et al.} showed theoretically that  counter-superflow is unstable when the relative velocity exceeds some critical value by linear analysis\cite{Law}.
Very recently Hamner {\it et al.} study experimentally and numerically shocks and dark-bright solitons in counterflowing BECs \cite{Hamner10}.  
Our work focuses on the drastic state of QT and the scenario of the instability towards QT. 
\begin{figure}
\centering
 \includegraphics[width=6cm]{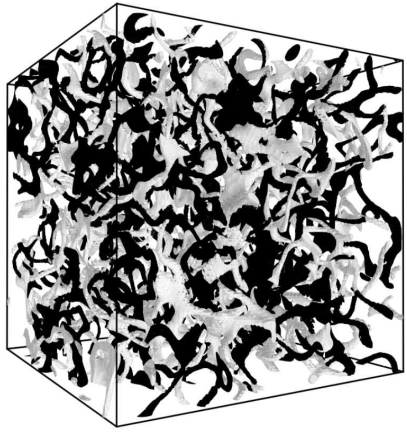}
 \caption{
  QT in two-component BECs.
  This shows the surface of the equal density, $n=0.1$.
  The black one and the gray one are the vortices of component 1 and component 2 .
 }
\end{figure}%

\section{Instability of counter-superflow}

 We consider binary mixture of BECs described by the condensate wave functions $\Psi _j=\sqrt{n_j}e^{i\phi _j}$ in the mean-field approximation at $T=0$ K, where the index $j$ refers to each component $j$ ($j=1,2$).
 The wave functions $\Psi _{j}$ are governed by the coupled Gross-Pitaevskii (GP) equations \cite{Pethick}
 \begin{equation}
 i \hbar\frac{\partial}{\partial t}\Psi _{1}= - \frac{{\hbar}^{2}}{2m_{1}}\nabla^2\Psi _{1}+V({\bm r})\Psi _{1}+g_{11}|\Psi _{1}|^{2}\Psi _{1}+g_{12}|\Psi _{2}|^{2}\Psi _{1},
 \end{equation}
 \begin{equation}
 i \hbar\frac{\partial}{\partial t}\Psi _{2}= - \frac{{\hbar}^{2}}{2m_{2}}\nabla^2\Psi _{2}+V({\bm r})\Psi _{2}+g_{22}|\Psi _{2}|^{2}\Psi _{2}+g_{12}|\Psi _{1}|^{2}\Psi _{2},
 \end{equation}
 where $m_{j}$ is particle mass associated with the species, $g_{jj}$ is intracomponent interaction and $g_{12}$ is intercomponent  interaction.
 Our analysis satisfies the condition $g_{11}g_{22}>g_{12}^2$ that two condensates are miscible.
 Following the parameters in the experiments\cite{Hamner10}, we consider the case that particle mass $m_{j}$ and interaction coefficient $g_{jj}$ of two condensates are equal, namely $m_{11}=m_{22}=m$ and $g_{11}=g_{22}=g$.

 We can obtain the critical relative velocity in counter-superflow from dynamical instability.
 In a stationary state, the wave functions are written as $\Psi _j^0=\sqrt{n_j^0}e^{i{\bm k}_{j}\cdot{\bm r}}$, where superfluid velocity is ${\bm V}_{j}=\frac{\hbar}{m}\nabla \phi _{j} = \frac{\hbar}{m}{\bm k}_{j}$.
We assume $n_{1}^0=n_{2}^0=n^0$.
The disturbance  $\delta \Psi _{j}$ around $\Psi _{j}^0$ is represented by
\begin{equation}
\delta \Psi _j =e^{i({\bm k}_{j}\cdot {\bm r}-\mu_{j} t)} \bigl\{ u_{j}e^{i({\bm q}\cdot{\bm r}-\omega t)}-v_{j}^*e^{-i({\bm q}\cdot{\bm r}-\omega t)}\bigr\}.
 \end{equation}
 Inserting these formulations into the GP-equation yields the Bogoliubov-de Gennes equation.
Solving the eigenvalue problem, the frequencies $\omega$ are found to obey
 \begin{equation}
(\hbar\omega-{\bm V}_{c}\cdot{\bm P})^2 =\epsilon^2+2\epsilon gn^0+\frac{1}{4}|{\bm V}_{R}\cdot {\bm P}|^2 \pm \sqrt{(\epsilon^2+2\epsilon gn^0)|{\bm V}_{R}\cdot {\bm P}|^2+4\epsilon^2 g^2 n^{02}\gamma^2}.
 \end{equation}

 Here ${\bm V}_{c}=({\bm V}_{1} + {\bm V}_{2})/2$ is the velocity of the center of mass, ${\bm V}_{R}={\bm V}_{1} - {\bm V}_{2}$ is the relative velocity, ${\bm P} = \hbar {\bm q}$ is the momentum of excitation, $\epsilon = \frac{{\bm P}^2}{2m}$ is the energy of excitation, and $\gamma = g_{12}/g$.
When an imaginary part of a frequency is nonzero,{\it i.e.} $(\hbar \omega-{\bm V}_{c}\cdot {\bm P})^2<0$, the system is dynamically unstable.
 Then we obtain the condition of instability of counter-superflow is
 \begin{equation}
4\Bigl(\frac{{\bm P}^2}{4m^2}+\frac{gn^0}{m}-\frac{gn^0}{m}\gamma\Bigr)<({\bm V}_{R}\cdot\hat{{\bm P}})^2<4\Bigl(\frac{{\bm P}^2}{4m^2}+\frac{gn^0}{m}+\frac{gn^0}{m}\gamma\Bigr).
 \end{equation}
\begin{figure}
 \includegraphics[width=5cm]{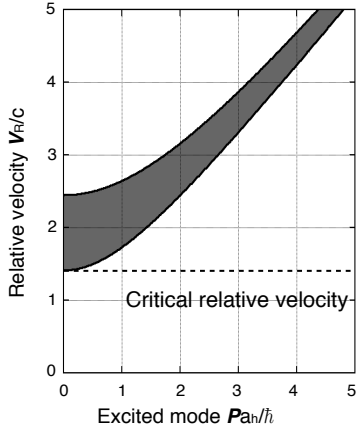}
 \caption{
 Unstable region of counter-superflow.
 The horizontal axis is the excited mode ${\bm P}$ normalized by $\hbar/a_{h}$ and the vertical axis is the relative velocity ${\bm V}_{R}$ normalized by sound velocity $c=\sqrt{gn^0/m}$.
 The gray zone shows Eq. (5).
 }
\end{figure}
Figure 2 shows this condition with $\gamma=0.5$.
The horizontal axis is the excited mode ${\bm P}$ and the vertical axis is the relative velocity ${\bm V}_{R}$.
The gray zone shows that counter-superflow is unstable and the ${\bm P}=0$ mode becomes unstable first with increasing the relative velocity ${\bm V}_{R}$.
Hence the critical relative velocity ${\bm V}_{{\rm critical}}$ is $2\sqrt{\frac{gn^0}{m}-\frac{gn^0}{m}\gamma}$.

 The dynamical instability should induce the exchange of momentum between the two components to reduce the relative motion of the condensates.
Here we may define the momentum of the component $j$  as ${\bm J}_{j}=\int d{\bm r} \frac{\hbar}{2i}(\Psi _{j}^*\nabla\Psi _{j} - \Psi _{j}\nabla \Psi _{j}^*)$.
In a stationary counter-superflow state with $\Psi_j=\sqrt{n_j} \exp(i\frac{m{\bm V}_j\cdot {\bm x}}{\hbar})$, we have ${\bm J}^0_j/V=mn_j{\bm V}_j$, where we use the volume $V$ of the system.
Let us suppose that a single unstable mode appears in the counter-superflow state.
Then the change $\delta {\bm J}_j$ in the momentum ${\bm J}_j$ due to the presence of the mode is written as $\delta {\bm J}_j/V=\hbar {\bm q}(|u_j|^2-|v_j|^2)$, where ${\bf q}$ is the wave number of the unstable mode and the particle number $N_j=\int dV|\Psi_j|^2$ is conserved. 
Since the total momentum $J_1+J_2$ must be conserved in this system, we require the condition $\delta {\bm J}_1+\delta {\bm J}_2=0$ and $\delta {\bm J}_j \neq 0$ to exchange momentum between the two condensates.
This condition is reduced to $|u_1|^2-|v_1|^2=-|u_2|^2+|v_2|^2\neq 0$, which is satisfied for the unstable modes of the counter-superflow instability.
Thus the momentum exchange is accelerated as the unstable modes are amplified by the counter-superflow instability.
 
\section{The dynamics of counter-superflow in a uniform system}

As described in Sec. 1, QT appears in counter-superflow.
We discuss the mechanism of transition to QT by investigating the dynamics of 1D and 2D systems.

 Firstly, we investigate the dynamics of 1D systems (Fig. 3).
Figure 3 (a) shows the time evolution of densities of two components.
 The horizontal axis is $x$ coordinate and the vertical axis is time.
The left of Fig. 3 (a) shows component 1 flowing from left to right and the right shows component 2 flowing oppositely, and the relative velocity is bigger than ${\bm V}_{{\rm critical}}$.
According to Fig. 2, when ${\bm V}_{R}=1.39{\bm V}_{{\rm critical}}$, the modes of $|{\bm P}|<1.36$ are unstable.
The density is initially uniform, but the waves of $|{\bm P}|<1.36$ are excited in both components by the dynamical instability.
Through the nonlinearity the waves grow into dark solitons, which travel without decaying.
The larger the magnitude of the imaginary frequency is, the less time it takes for the solitons to appear.
The solitons enjoy the repulsive interaction with other solitons.
The interaction randomizes the direction of propagation of solitons.
Two BECs exchange momentum, thus momentum decaying as shown in Fig. 3 (b).
Eventually, in each component, the number of solitons becomes almost constant and the numbers of solitons propagating to the right and to the left become almost equal, so soliton turbulence appears \cite{CGL}.
\begin{figure}
\centering
 \includegraphics[width=10cm]{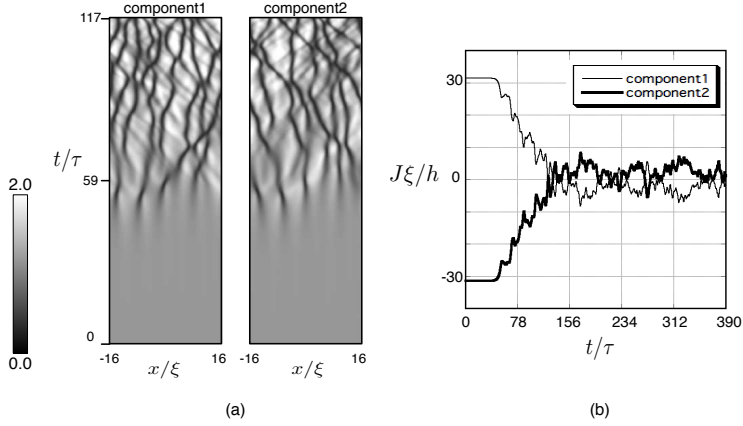}
 \caption{
(a) Time evolution of density of both components in a 1D system.
 The horizontal axis is ${\bm x}$ coordinate normalized by the coherence length $\xi$ and the vertical axis is time $t$ normalized by $\tau = \hbar/gn^{0}$.
 The parameters are $n=1.0$ and $|{\bm V}_{R}|/{\bm V}_{{\rm critical}}=1.39$ at $t=0$, and $\gamma = 0.5$.
 (b) Time evolution of momentum.
 The horizontal axis is time $t$ normalized by $\tau$ and the vertical axis is momentum ${\bm J}$ normalized by $\xi/h$.
 }
\end{figure}%

 Secondly, we investigate the dynamics in 2D systems.
 As 1D systems, unstable waves are excited and grow into dark solitons.
 Dark solitons are generally unstable in systems except for 1D systems, so they decay into vortex pairs by snake instability\cite{Anderson}.
The details will be reported elsewhere.

 A 3D system shows the rich dynamics including the transition to QT as well as the soliton dynamics like what occurs in 1D and 2D systems.
 Waves are excited and grow into dark solitons.
 The solitons decay into vortices or vortex rings by snake instability in 3D systems.
 These vortices repeat reconnections to develop to QT.
 Then both components are quantum turbulent, where two species of QT are interactive (Fig. 4).
\begin{figure}
\centering
 \includegraphics[width=12cm]{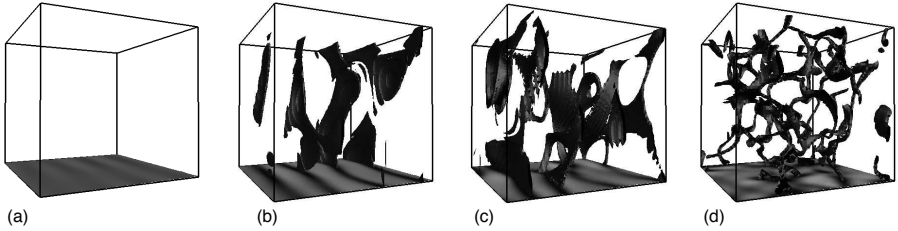}
 \caption{
 Surface of the equal density $n=0.1$ of one component in 3D two-component QT.
 The parameters are $n=1.0$, $|{\bm V}_{R}|/|{\bm V}_{{\rm critical}}|=1.66$ at $t=0$, and $\gamma = 0.5$.
(a) $t/\tau =0$. Initial state.
(b) $t/\tau =39$. Waves are excited by the dynamical instability and grow into dark solitons.
(c) $t/\tau =43$. Solitons decay into vortices by snake instability.
(d) $t/\tau =51$. QT appears.
 }
\end{figure}%

\section{Dynamics of counter-superflow in a 2D trapped system}
 We propose how to create two-component QT.
 We address two species of BECs with different internal degrees of freedom in a harmonic trap.
 Let two BECs displaced oppositely from the center of the potential by applying a magnetic field gradient, then reversing the direction of the field gradient and have two BECs collide.
 QT should appear like a uniform system, but it is difficult to study numerically in a 3D trapped system.
 Hence we address 2D trapped systems as the first trial.
 We suppose that two cigar-shaped $^{87}$Rb BECs collide with each other.
 In the initial state BECs are separated by a magnetic field gradient (Fig. 5 (a)).
 Reversing the direction of the field gradient, BECs move oppositely and collide each other.
 Waves are excited and grow into solitons like a uniform system (Fig. 5 (b), (c))\cite{Hamner10}.
 Eventually, solitons decay into vortex pairs (Fig. 5 (d)).
 Then BECs exchange momentum. 
 Thus we obtain the same dynamics as in a uniform system. 
 We expect that QT should appear in a 3D system.
\begin{figure}
\centering
 \includegraphics[width=12cm]{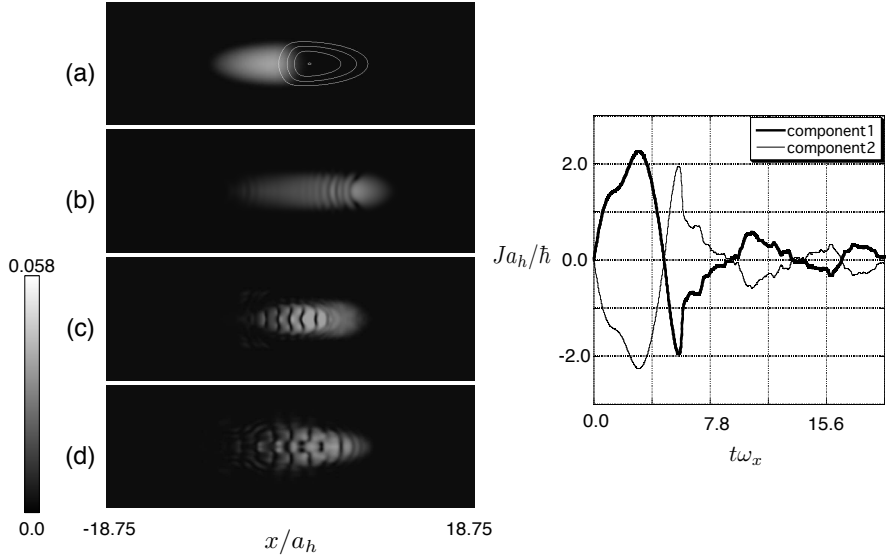}
 \caption{
 (Left) Time evolution of the density in a 2D trapped system.
 Parameters are $m=1.5\times 10^{-25}$kg, $a=5.00$nm, and two compoents have same particle numbers $N=1\times 10^5$.
 The trap frequencies are $\omega _{x}=2\pi\times 141$, $\omega _{y}=2\pi\times 445$ and $\omega _{z}$ is much smaller than $\omega _{x}$ and $\omega _{y}$.
 The coordinates ${\bm x}$ and ${\bm y}$ are normalized by the trap length $a_{h} = \sqrt{\hbar/m\omega _{x}}$.
 The value of density is shown as the color bar.
 (a) $t\omega _{x} =0.0$. Initial state.
 (b) $t\omega _{x}=3.9$. Waves are excited.
 (c) $t\omega _{x}=6.0$. Solitons appear.
 (d) $t\omega _{x}=6.4$. Solitons decay into vortex pairs.
 (Right) The time evolution of momentum of two components.
 }
\end{figure}%

\section{Summary}
 We studied the nonlinear dynamics of the counter-superflow instability in two miscible BECs.
 The instability induces the momentum exchange between the two condensates and the soliton formulations.
 Then the solitons decay into quantized vortices and vortices grow into the quantum turbulence of two superfluids.
 We found that the similar dynamics can occur in a 2D trapped system.
Therefore, we expect that QT should appear in a 3D trapped system too.
We believe that the counter-superflow QT offers a new avenue for study of turbulence.
The detail will be reported soon elsewhere.

\begin{acknowledgements}
 M.T. acknowledges the support of a Grant-in-Aid for Scientific Research from JSPS (Grant No. 21340104).

\end{acknowledgements}

\end{document}